# PSNR and Jitter Analysis of Routing Protocols for Video Streaming in Sparse MANET Networks, using NS2 and the Evalvid Framework

Sabrina Nefti
Dept. of Computer Science
University Batna 2
Algeria

Maamar Sedrati
Dept. of Computer Science
University Batna 2
Algeria

*Abstract*—Advances in multimedia and ad-hoc networking have urged a wealth of research in multimedia delivery over ad-hoc networks. This comes as no surprise, as those networks are versatile and beneficial to a plethora of applications where the use of fully wired network has proved intricate if not impossible, such as prompt formation of networks during conferences, disaster relief in case of flood and earthquake, and also in war activities. It this paper, we aim to investigate the combined impact of network sparsity and network node density on the Peak Signal Noise to Ratio (PSNR) and jitter performance of proactive and reactive routing protocols in ad-hoc networks. We also shed light onto the combined effect of mobility and sparsity on the performance of these protocols. We validate our results through the use of an integrated Simulator-Evaluator environment consisting of the Network Simulator NS2, and the Video Evaluation Framework Evalvid.

## I. INTRODUCTION

The transmission of multimedia objects over Mobile Ad-hoc Networks (MANET) network has become the need of the day due to critical applications that rely on such networks such as the transmission of important images and videos in emergency situations. However, this task presents two main complexities. The first aspect of intricacy lies in the nature of MANET: their mobile and distributed, interference and multi-hop communication [1], [2]. Since MANETs do not rely on pre-existing infrastructure, data is transmitted through multi-hop routing [3]. This collective effort in data transmission requires that each node acts as a router too. Thus, it comes as no surprise that the provision of QoS over such networks can prove extremely difficult. The second aspect of intricacy lies within the nature of multimedia objects, specifically video files which are not only bandwidth-hungry but also highly-demanding in terms of Quality of Service (QoS). Effective multimedia transmission dictates minimal delay and in-order receipt of packets [1]. Therefore, it has become imperative to determine routing protocols that can not only fulfil those QoS criteria but are also able to maintain such performance while varying the network topology in terms of sparsity and mobility. Multimedia transmission may prove particularly challenging in sparse MANETs whereby disconnections become more and more frequent due to low network node density [3].

It this paper, we investigate the PSNR performance of proactive and reactive routing protocols for video streaming of bandwidth-hungry multimedia video files over sparse MANET networks. We also explore the combined effect of mobility and sparsity on the PSNR performance. To this end, we use the NS2 Network Simulator and the Evalvid Framework tool in order to test a renowned protocol of each family, namely AODV (reactive), and DSDV (proactive).

The remainder of the paper is organized as follows: Section 2 explores the previous work performed in this field. A brief description of the System Model adopted in our work is presented in Section 3. Section 4 justifies our choice of the simulation and evaluation tools used. Next, a detailed work approach is described along with simulation configuration in Section 5. Results are presented and analyzed in Section 6. Finally, conclusions and future work recommendations are provided in Section 7.

## II. RELATED WORK

Various comparative studies have been carried out between proactive and reactive protocols [4], [5]. In [5], the QoS metrics used for comparison are media access delay, network load and throughput. The study in [6] is similar to [5] with the addition of retransmission attempts metric. However, the afore-mentioned studies have not taken into consideration the augmented challenges dictated by the transmission of quality-demanding multimedia objects. A more specific analysis of routing protocols for video streaming was undertaken in [7], whereby two network structures (25 nodes and 81 nodes) were simulated using OPNET in order to assess QoS parameters such as throughput, wireless LAN delay, end-to-end delay and packet delay variation. In our work, we investigate the performance of MANET routing protocols in video streaming on the basis of Peak to Signal Noise Ratio Video Quality Model, also called $VQM_p$ [8] as well as jitter.

### III. SYSTEM MODEL

Fig. 1 shows the overall System Model: initially a video file in the raw YUV format is converted into a MPEG4 file. This latter is fed into Evalvid to generate a video trace file which is the actual video object that is sent over a simulated transmission in NS2. The received video file received at the receiver node in the simulation environment is fed back into Evalvid which generates the PSNR quality model amongst various other QoS metrics [9].

The PSNR quality model is considered as one of the most widespread models in assessing video quality in an objective manner, as it was developed specifically to emulate the quality impression of the Human Visual System (HVS) [11]. Furthermore, this model is a derivative of the notorious Signal to Noise Ratio (SNR). However, while SNR compares the signal energy to the error energy, PSNR compares the maximum possible signal energy to the noise energy. This subtle difference has shown to yield higher correlation with the subjective quality perception than the conventional SNR [12]. The following equation is the definition of the PSNR between the luminance component Y of source image S and destination image D [11]:

$$PSNR(n)_{dB} = 20 \log_{10} \left( \frac{V_{peak}}{\sqrt{\frac{1}{N_{col} N_{row}} \sum_{i=0}^{N_{col}} \sum_{j=0}^{N_{row}} [Y_s(n,i,j) - Y_D(n,i,j)]^2}} \right) \quad (1)$$

$$V_{peak} = 2^k - 1 \quad k = \text{number of bits per pixel} \quad (2)$$

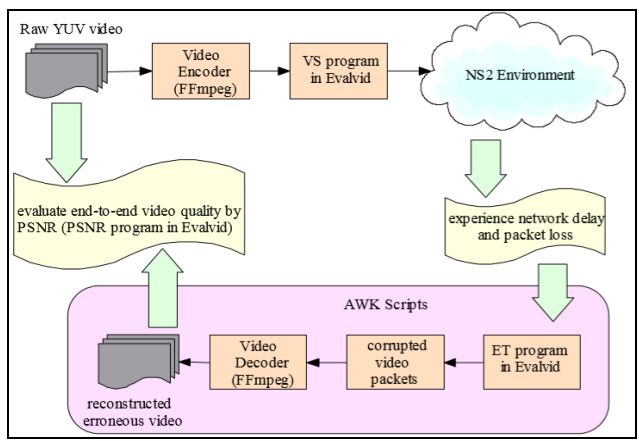

Fig. 1: System Model: Evalvid Framework integration with NS2 [10]

### IV. SIMULATION AND EVALUATION TOOLS

A survey of simulation tools for MANET has revealed that the NS2 simulator is one of the most widespread tools for MANET network simulations with a usage percentage of 43.8%, largely outperforming its direct competitor namely the bespoke tools (27.3%), as can be depicted in Fig. 2 [13]. Based on this finding, we opted for NS2 in our work.

Two main video quality evaluation tools were identified in the literature, namely the MSU tool used to extract the Structural SIMilarity (SSIM) index [14] and the Evalvid Framework, which evaluates the quality of videos transmitted over real networks [11]. The advantage that Evalvid presents compared to MSU is that it can be integrated into NS2, and therefore it is possible to use Evalvid to evaluate the quality of a video object transmitted in an NS2 simulation environment [12], hence the reason for which we selected Evalvid Framework in our work.

#### A. The Evalvid Tools

The Evalvid Framework consists of three main tools [10]. In the following tables, we summarize the functionality and the parameters of these tools based on their execution in the Windows command line tool (cmd).

TABLE 1: FUNCTIONALITY AND PARAMETERS OF THE "MP4TRACE" TOOL

| Tool | mp4trace |
|---|---|
| Functionality | Converts the MPEG4 video file to be transmitted into a video trace file. This trace file is then sent by the source node in the NS2 simulation environment to the receiver node. |
| Usage | mp4trace [options] <file 1 > <file 2> |
| Parameters | Options:<br>**-[p/f]** packet or frame mode<br>**-s** host port: sends the RTP packets to specified host and UDP port<br>**<file 1>** the MPEG4 video file to be transmitted<br>**<file 2>** the generated trace file generated by the tool |

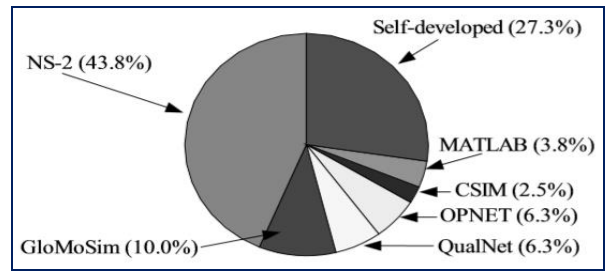

Fig. 2: Utilization percentages of Simulation Software for MANET [13]

TABLE 2: FUNCTIONALITY AND PARAMETERS OF THE "ETMP4" TOOL

| Tool | etmp4 |
|---|---|
| Functionality | 1. The video trace file "**st**" generated by *mp4trace* (Table1) is fed into NS2 and two files are generated from the simulation: the sender's frame transmission times file "**sd**", and the |

| | receiver's frame reception times file "**rd**". The *etmp4* function generates the mpeg4 video file "**out**" that was received at the receiver based on the "**st**", "**sd**" and "**rd**" files as well as the original transmitted video file.<br>2. This tool also generates QoS metrics such as loss rate, debit etc... |
|---|---|
| Usage | etmp4 -[p\|f\|F] -[0\|x] [-c] <sd> <rd> <st> <in> <out> |
| Parameters | **-[p\|f\|F]** packet, frame or complete frame mode<br>**-[0\|x]** fill lost section with 0 or truncate<br>**[-c]** use cumulative jitter in case of asynchronous clocks<br>**<sd>** tcpdump sender<br>**<rd>** tcpdump receiver<br>**<st>** trace-file sender<br>**<in>** transmitted video (original mp4)<br>**<out>** base name of output file<br>**[PoB]** optional Play-out buffer size [ms] |

TABLE 3: FUNCTIONALITY AND PARAMETERS OF THE "PSNR" TOOL

| Tool | Psnr |
|---|---|
| Functionality | This function generates the PSNR metric image by image according to the afore-mentioned formulae. |
| Usage | psnr x y <YUV format> <src.yuv> <dst.yuv> > |
| Parameters | **x** frame width<br>**y** frame height<br>**YUV format:** 420, 422, etc..<br>**src.yuv:** source video<br>**dst.yuv:** distorted video |

### B. The Integration of Evalvid into NS2

In order to integrate the Evalvid Tool in NS2, two main types of amendments are required:

- Modifications applied in the NS2 header files and Makefile, these are explained in detail in [10].

- The addition of new C++ classes into the NS2 core code [15][10]: Those classes augment the NS2 environment with objects that have the capability to transmit video trace files (generated by the *mp4trace* tool described in Table 1). We analyzed the code of these classes and presented our understanding of their augmented video capabilities in Table 4.

TABLE 4: THE CONTRIBUTION OF THE NEW CLASSES

| New Class | The Contribution of the New Class in the Video Streaming Simulation |
|---|---|
| myUDP | *myUDP* class extends the NS2 Agent Class (and hence has access to the latter's functions, thanks to the inheritance principle). An object of *myUDP* represents the UDP transport protocol and has two main video capabilities:<br>1. The function «*attach-agent*», attaches the *myUDP* object to a *myEvalvid* object. Thanks to this attachment, the *myUDP* object can extract the video trace file (to be transmitted) from the *myEvalvid* object (to which it is attached).<br>2. The function "*set_filename*", passes to the object *myUDP* a file pointer in which it can record the transmission time of each transmitted frame (or packet). |
| MyEvalvid_Sink | Similar to the above, the *MyEvalvid_Sink* class extends the NS2 Agent Class. The main two functionalities that contribute to the video transmission process over NS2 are as follows:<br>1. The function «*connect*» connects the *MyUDP* object to the *MyEvalvid_Sink* object. This connection renders possible the transmission of the video trace file (attached to the *myUDP* sender object) to the receiver object *MyEvalvid_Sink*.<br>2. The function « *set_filename* » allows the *MyEvalvid_Sink* object to record the reception time of each received frame (or packet). |
| myEvalvid | *myEvalvid* Class extends the NS2 Traffic Class. An object of type *myEvalvid* represents a traffic source of type video and has two main capabilities:<br>1. Thanks to the function « *attach-tracefile*», the *myEvalvid* object can be attached to an object of type *Trace* (an inherent NS2 Class). The *Trace* object, in turn, can be attached to a video trace file (generated by *mp4trace* tool) thanks to the function "*filename*". This double attachment enables the *myEvalvid* object to be attached to the video trace file (to be transmitted).<br>2. Thanks to the function « *attach-agent*», the *myUDP* object can be attached to the traffic source object *myEvalvid* which encompasses the video trace file (see point 1 above). This attachment enables the transport object *myUDP* to transport the video trace file (as this data can be obtained from the traffic source object *myEvalvid* attached to the transport object *myUDP*). |

### V. WORK APPROACH

Our work approach is summarized in four main steps as shown in Fig. 3. Each step in explained separately in the following sub-sections.

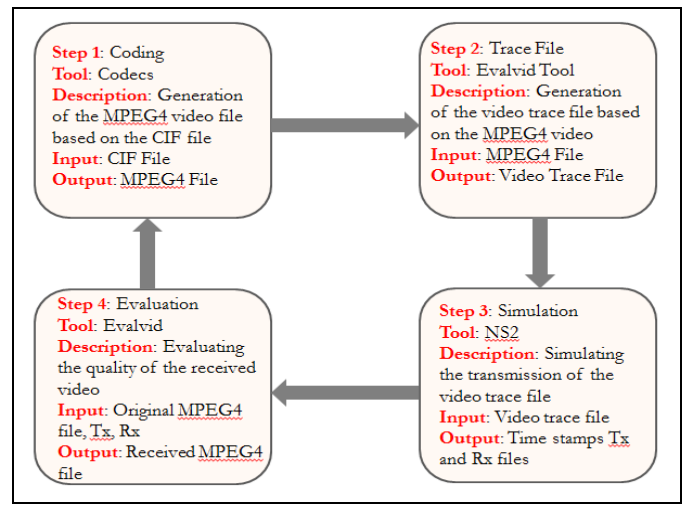

Fig. 3: Work Steps: (1) Coding (2) Trace File Generation (3) NS2 Simulation (4) Evaluation in Evalvid

A. *The Coding Process*

The type of video files used in our simulations are the H.261 standard Common Interface Format (CIF) with 352 × 288 resolution, as this format is commonly used in video teleconferencing, which is one of the important applications of MANET. Before a CIF file could be used for simulation purposes, it is first encoded into MPEG format, one of industrial standards widely used in video streaming over the internet [16]. This coding occurs in three stages [10]:

- First the CIF file is converted into a YUV file, a video data format which takes into account human perception. This is performed by the ffmpeg codec [17]. The command line is as follows:

    *ffmpeg -i CIF_File Original_Yuv_File.yuv*

- The Yuv file resulting from the above is converted MP4V format with the xvi codec [17]. This format is considered as the intermediary raw format of MP4:

    *xvid_encraw -i Original_Yuv_File.yuv - w 352 -h 288 -framerate 30 -max_key_interval 30 -o Original_MV4_File.m4v*

- Finally, the M4V file (*Original_MV4_File.m4v*) is coded into a MPEG4 file (*Original_MP4_File.mp4*) with the MP4Box codec [17], with the following command line:

    *MP4Box -hint -mtu 1024 -fps 30 –add Original_MV4_File.m4v Original_MP4_File.mp4*

B. *Video Trace File Generation*

The second step consists of generating a video trace file from the original MPEG4 video file using the *mp4trace* tool described in Section 4.1. The video trace file contains the frame number, type and size and the number of segments in case of frame segmentation [11]. The Evalvid tool was originally designed to evaluate real video transmissions, hence the reason why *mp4trace* tool specifies the destination URL and port number. However, for the sake of our work, this tool is executed with an arbitrary IP and Port number, as the sole aim of this execution is the generation of the video trace file and not its actual transmission over the internet:

*mp4trace -f -s 192.168.0.2 12346 Original_MP4_File .mp4 > eval_trace_file*

C. *Simulation in NS2*

In the first set of simulations, we aim to investigate the combined impact of the network node density and network sparsity on the PSNR performance of AODV and DSDV. To this end, we designed a matrix-based network topology whereby nodes are symmetrically placed in a matrix with equal horizontal and vertical distances from each other. By network sparsity, we mean how distanced or close the network nodes are, whereby the distance between each neighboring nodes in the matrix topology is referred to by Distance (D). By network node density, we mean the number of nodes that the network consists of.

We begin our simulation with a network with nodes that are close to each other (as opposed to sparse network), and gradually disperse it, by equally augmenting the vertical and horizontal distance between each two neighboring nodes. The distances considered are 20m, 50m, 100m, and 150m. In order to test the combined effect of network node density and its sparsity, we test each network sparsity model with several network nodes density ranging from 4, 9, 16, 25, 36, 49 and 64 nodes. Table 5 shows our simulation configuration.

In the second set of simulations, we test the impact of mobility that results in sparsity on the PSNR performance of both AODV and DSDV. To this end, we start the simulation with a network in which the nodes are closely distanced from each other (D=20m), and which move outward with a constant speed; in order to form a sparse matrix (D=150m), as can be perceived in Fig. 4. We refer to this scenario in the remainder of our paper as *Outward Mobility*.

In the third and last set of simulations, we test the effect of mobility that results in a network with closely distanced nodes on the PSNR performance of both protocols. To this end, we commence the simulation with a sparse network (D=150 m), in which nodes move inward with a constant speed to form a network with closely distanced nodes (D=20m). We refer to this scenario in the rest of our paper as *Inward Mobility*.

TABLE 5: SIMULATION CONFIGURATION

| Simulation Parameter | Configuration |
|---|---|
| Propagation Model | TwoRayGround |
| MAC | 802.11 |
| Routing Protocols | AODV, DSDV |
| Placement of Nodes | Matrix-based placement with equal vertical and horizontal distance between nodes. This distance varies from: 20m, 50m, 100m, and 150m |
| Number of Node | 4, 9, 16, 25, 49, and 64 Arranged in matrices of: 2 x 2, 3 x 3, 4 x 4, 5 x 5, 6 x 6, 7 x 7, and 8 x 8 |
| Video File Frame Size | 2000 frames |

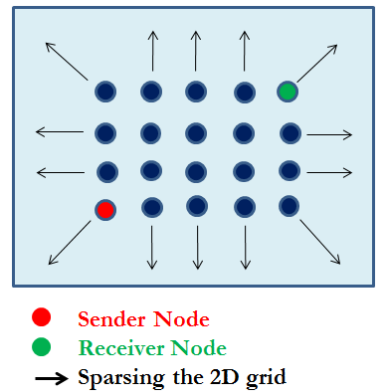

● Sender Node
● Receiver Node
→ Sparsing the 2D grid

Fig. 4: Outward Mobility: The nodes are dispersed as a result of mobility.

*D. Evaluation of the Video Quality*

We evaluate of the quality of the received video file in three stages [10] [11]:

- The generation of the received video file: The Evalvid tool etmp4 compares the sender's timestamp file (which contains the transmission time of each transmitted frame) against the receiver's timestamp file (which contains the reception time of each received frame). Through this comparison, and the original MP4 video and the video trace file, the tool reconstructs the received MP4 video. During this reconstruction process, the tool also measures the delay per frame, the frame loss rate, the instantaneous transmission as well as well as the reception debit. The corresponding command is as follows:

    *etmp -f -0 <s_time_trace> <r_time_trace> <video_trace_file> Original_Mp4_File.mp4 Received_MP4_File.mp4*

- The generation of the received Yuv file: using the ffmpeg codec and the received video (from the previous step). The required command is as follows:

    *ffmpeg -i Received_MP4_File.mp4 Received_Yuv_File.yuv*

- The generation of the PSNR metric: using the psnr Evalvid tool which compares the original and the received Yuv files in order to calculate the PSNR per frame. The corresponding command is as follows:

    *psnr 352 288 420 Original_Yuv_File.yuv Received_Yuv_File.yuv*

## VI. RESULTS AND ANALYSIS

*A. PSNR Performance Analysis*

The video used in all our simulations is based on "*Highway CIF*" [18]. Fig. 5 shows the PSNR performance of AODV when the network sparsity is fixed to D=100m and its density is increased. A moving average filter of a 100 frames width was used to smooth the results a clearer analysis.

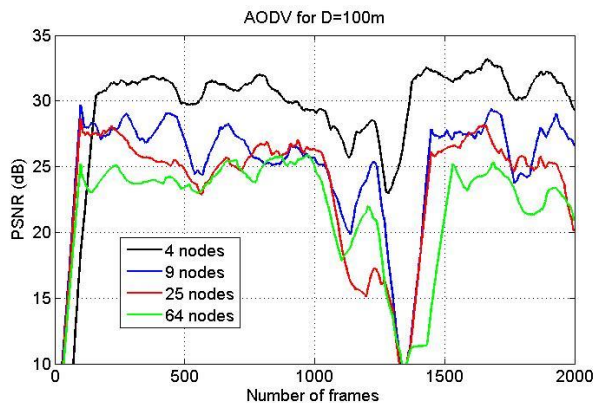

Fig. 5: PSNR Performance of AODV for various network densities (4, 9, 25, 64) when the distance is fixed to D=100.

The PSNR variation pattern is maintained across various network densities. The PSNR variation pattern is affected by the luminosity content of the frames that the transmitted video consists of. Since we are transmitting the same video file across varying topologies, the luminosity content of each frame of this video remains constant. We further verified this by investigating the two main drops in PSNR performance in Fig. 5. The first one occurred approximately at frame F=550, corresponding to the approximate video play time of T=21s. We observed that during this timeframe, the luminosity is decreased due to the appearance of an overtaking black car, as can be depicted in Fig. 6.

The second PSNR major drop occurred between frames F=1250 and F=1300, corresponding to the approximate video play times of T=41s and T=43s. During this timeframe a dark bridge first appears in the video and then the car passes under its shadow as shown in Fig. 7.

Therefore, the PSNR drop in the two cases of Fig. 6 and Fig. 7 can be justified by the fact that when the luminosity content of a frame decreases, the noise energy dominates over the peak signal energy, and hence degrading the PSNR.

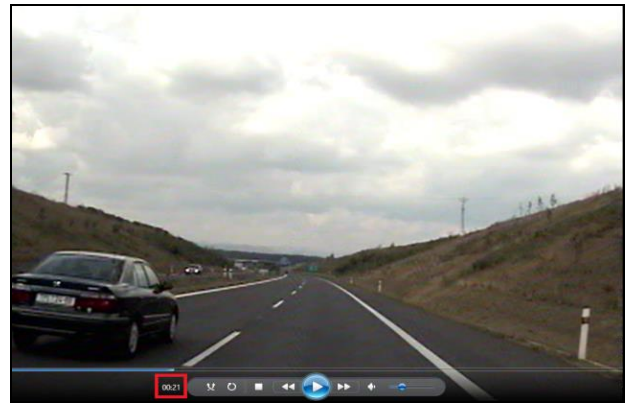

Fig. 6: Appearance of a black car at T=21s

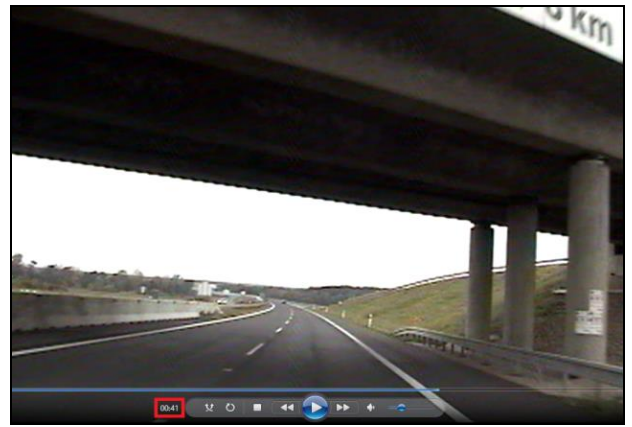

Fig. 7: Appearance of a bridge and its shadow at T=41s

It can also be noted from Fig. 5 that the PSNR performance of AODV degrades as the density of the network increases. The PSNR decreases by approximately 5dB between frames Frame=400 and Frame=600 when upsizing the network from N=4 nodes to N=9 nodes. However, this attenuation is less significant when the network is further upsized to N=25 nodes and N=64 nodes. This is due to the fact that the 4-nodes topology allows direct communication between the sender and the receiver nodes; however, as the network density is augmented to N=9 nodes, data is routed through intermediary nodes. In this case, multi-hopping decreases the PSNR performance significantly compared to direct sender receiver one-hop transmission.

It can also be seen that AODV registers a sharp rise in PSNR between frame F=0 and F=100 for the various network topologies, which can be justified by its fast convergence in low density networks [19].

Fig. 8 compares the PSNR performance of AODV and DSDV in two different topologies. A moving average filter of a 100 frames width was used to smooth the results for a clearer analysis. It is clear that AODV outperforms DSDV in small sparse networks (D=100m, N=4), with a PSNR difference ranging from 5dB to 23dB. When the distance between two neighboring nodes is halved from D = 100m to D=50m, the PSNR performance of both protocols increases by a range of 3dB to 10dB and also becomes smoother.This occurs despite the fact that the network density is quadrupled from N=4 to N=16 nodes.

In this scenario, it can be concluded that PSNR performance of both AODV and DSDV is better in high-density and low-sparsity networks than in it low-density and high-sparsity ones. In fact, extensive simulations demonstrated that when increasing the network sparsity, DSDV protocol is unable to deliver a video quality that is sufficient enough to extract the PSNR metric; Table 6 captures these cases.

In order to verify this finding, we reconstructed and played the video sent over a network topology of N = 64 nodes and D=100m in which DSDV was set as the routing protocol, and noted that it was significantly distorted for more than half of the video length as captured in Fig. 9.

*B. Jitter Performance Analysis*

Figure 10 shows the jitter performance of ADOV when varying the network density. It can be remarked that the jitter variation pattern remains similar when varying network densities.

Similar to PSNR, the jitter variation pattern is also dependent on the luminosity content of the frames transmitted. However, unlike PSNR, jitter performance improves when the frame luminosity content is low; as such frames carry less data content than frames with high luminosity, and hence enjoy better delay variation performance. Indeed, the jitter performance improves at approximate frames F=550 and F=1250 which have low luminosity content as discussed earlier, and shown in Figure 6 and Figure 7.

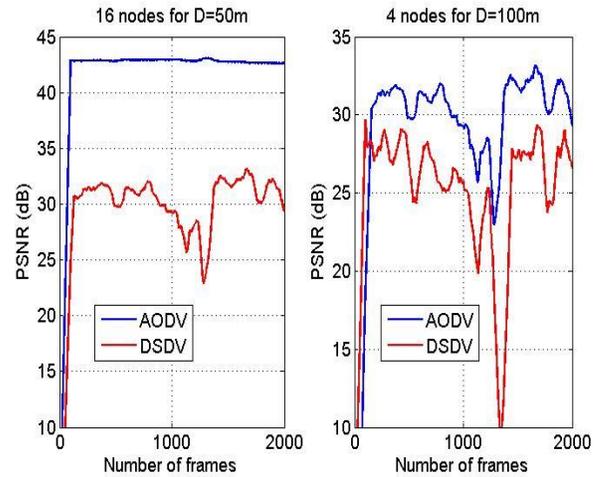

Fig. 8: Comparison of PSNR performance of AODV and DSDV: for two scenarios (D=50m, Nodes=16) and (D=100m, Nodes =4).

TABLE 6: CASES WHERE THE PSNR GENERATION WAS NOT POSSIBLE DUE TO INCREASED NETWORK SPARSITY (A: AODV, D: DSDV, Y: PSNR GENERATED, N: PSNR NOT GENERATED)

| No of Nodes | D=20 | | D= 50 | | D=100 | | D=150 | |
|---|---|---|---|---|---|---|---|---|
| | A | D | A | D | A | D | A | D |
| 4 | Y | Y | Y | Y | Y | Y | Y | Y |
| 9 | Y | Y | Y | Y | Y | **N** | Y | **N** |
| 16 | Y | Y | Y | Y | Y | **N** | Y | **N** |
| 25 | Y | Y | Y | **N** | Y | **N** | Y | **N** |
| 36 | Y | Y | Y | **N** | Y | **N** | Y | **N** |
| 49 | Y | Y | Y | **N** | Y | **N** | Y | **N** |
| 64 | Y | Y | Y | **N** | Y | **N** | Y | **N** |

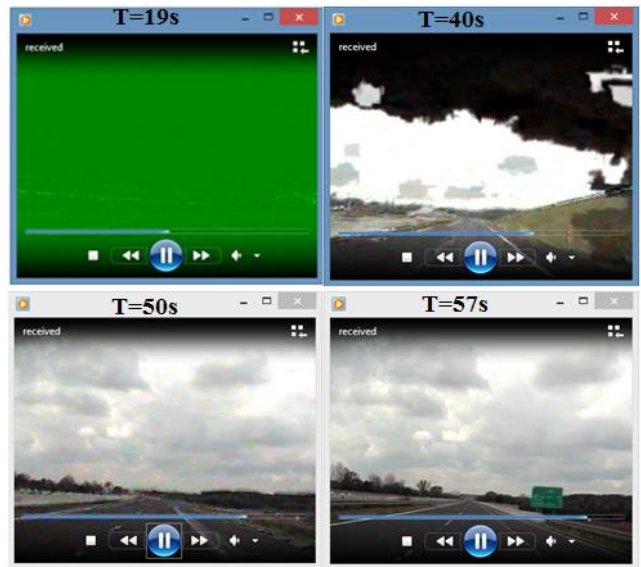

Fig. 9: screen captures of the video received across a network topology of D = 100m, N=64 nodes, DSDV protocol

Fig. 10 also reveals that as we upsize the network, the jitter value is slightly degraded with the worst jitter being registered for N=64 nodes and the overall jitter ranging from approximately -0.1s to 0.05s. This suggests that AODV's jitter performance shows a certain degree of resilience to increasing the network density.

Fig. 11 depicts the jitter performance of DSDV for the same network topologies as Fig. 10. As the network density is augmented, the jitter metric is significantly degraded, with the jitter ranging from -1.5s to -0.2s. Therefore, the DSDV jitter performance shows less resilience to network upsizing than AODV. This degradation is due to the delay variation incurred by the additional multi-hopping that takes place when the number of intermediary nodes through which data has to be routed increases.

In order to analyze the jitter performance more closely, we considered the various network topologies in Fig. 12. It is clear that AODV outperforms DSDV in all four scenarios, what is interesting to note though, is that for AODV, jitter varies between a small range -0.05s and 0s even when augmenting both the network density and sparsity (from N=16 to N=49 and from D=20m to D=100m). This suggests that AODV's jitter performance is robust to variations in both network density and sparsity.

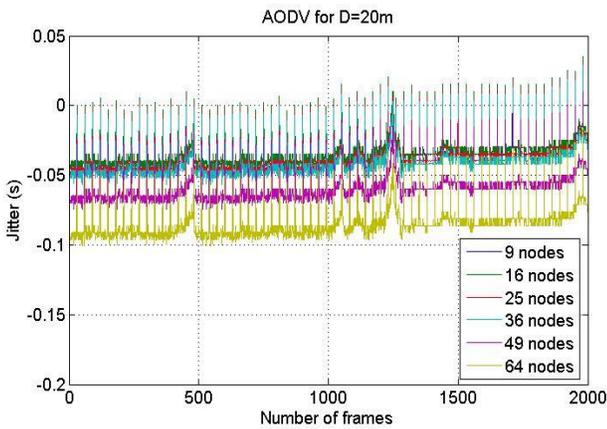

Fig. 10: Jitter performance of AODV with varying network densities (D=20m)

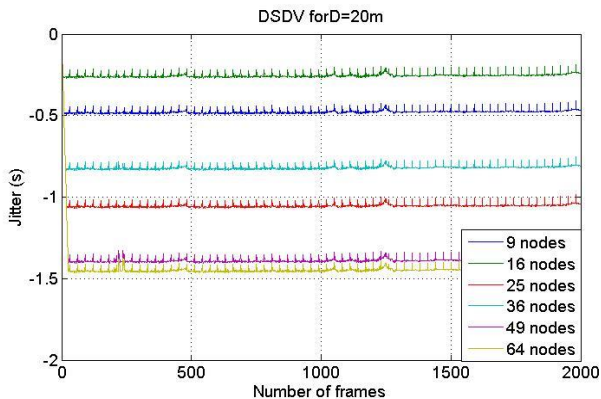

Fig. 11: Jitter performance of DSDV protocol with varying network densities (D=20m)

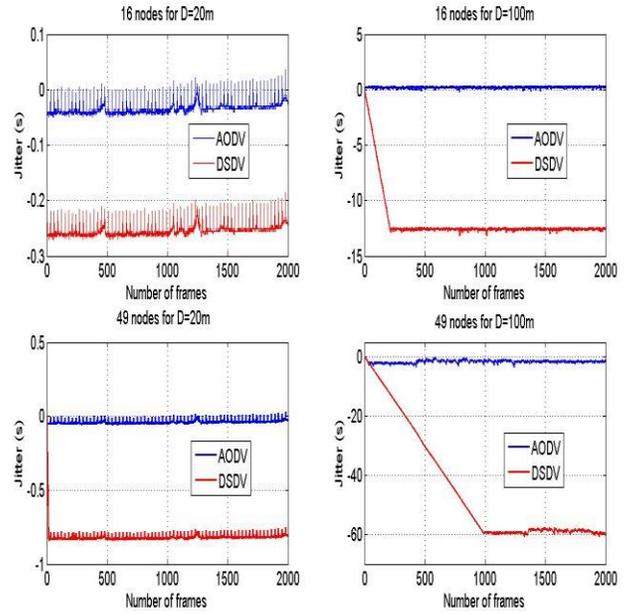

Fig. 12: Comparison of jitter performance for AODV and DSDV for various topologies

DSDV's jitter performance dropped from -0.25s to -0.8s when maintaining the network sparsity (D=20m) and increasing its nodes density (from N=16 to N=49). However, the DSDV's jitter degraded significantly from -0.25s to -12s when the network density was fixed to N=16 and the network sparsity was augmented from D=20m to D=100m. Furthermore, when the density was fixed to N=49, an even more considerable drop in DSDV's jitter performance occurred (from -0.8s to -60s) when the sparsity increased from D=20m to D=100m.

This suggests that AODV's jitter performance shows a much better resilience to the increase in network sparsity and density than DSDV. Furthermore, DSDV's jitter performance is much more resilient to network density than it is to network sparsity, as the results showed that increasing the sparsity of the network degraded the jitter performance more significantly than increasing its density. It can also be noted that as the sparsity increased from D=20m to D=100m, the DSDV's jitter falls sharply in the initial F=250 frames and F=1000 frames respectively, this is due to the DSDV's slow convergence in low density networks [19] as well as increasing the network sparsity.

C. *Mobility: PSNR Performance Analysis*

Fig. 13 depicts the PSNR performance of a 5x5 matrix in the *Outward Mobility* scenario, i.e. the network's initial sparsity is set to D=20m, and it increases as the nodes move outward with a constant speed to eventually form a sparse matrix of D=150m.

It can be observed that AODV's PSNR performance significantly outperforms DSDV's during the first F=800 frames. This is due to DSDV's slow convergence caused by periodic updates of routing tables. This is combined with the fact that the network is being dispersed by the nodes movement and hence the update of routing tables between two one-hop neighbours takes longer as the neighbouring nodes are moving away from each other.

However, from frame F=800 onwards, DSDV registers close levels of PSNR compared to AODV. What is interesting to note, is that as the sparsity increases toward the end of the simulation, DSDV's PSNR performance outperforms AODV's. As the mobile nodes become more distant from each other, AODV's reactive route discovery process becomes less efficient, hence reducing the signal strength of each frame compared to its noise content. However, for DSDV, once the routing tables' updates have taken place, DSDV's pre-calculated routes allow for a faster route discovery, hence why DSDV demonstrates a better PSNR performance towards the end of the simulation despite the increasing sparsity of the network.

The opposite scenario is where the network's initial sparsity is set to D=150m and it gradually shrinks to D=20m with the nodes moving inward with a constant speed. Due to the fact that DSDV PSNR metric cannot be extracted when the network is initially very sparse (Table 6), we could not analyse DSDV's PSNR performance in the *Inward Mobility* scenario.

Fig. 14 compares AODV's PSNR performance in *Outward Mobility* against *Inward Mobility*. It can be observed that for approximately the initial 1300 frame, *Outward Mobility* outperforms *Inward Mobility* in terms of PSNR metric. In *Outward Mobility*, PSNR is stronger initially as the nodes are closely positioned from each other, and gradually decreases as the network becomes sparser. This explains the reason why in *Inward Mobility*, the PSNR is weaker initially but eventually outperforms the *Outward Mobility*, as the nodes get closer to each other towards the end of the simulation. Hence, it can be concluded that not only PSNR performance is affected by mobility but is also sensitive to the effect that this mobility has on the network, i.e. whether the mobility results in a sparse network or a closely populated one.

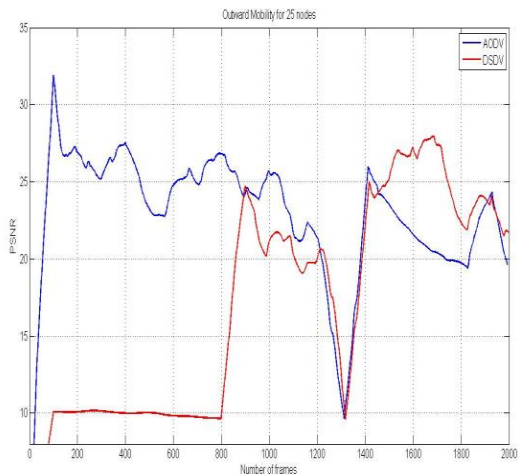

Fig. 13: The impact of Outward Mobility on PSNR for AODV and DSDV (N=25 nodes)

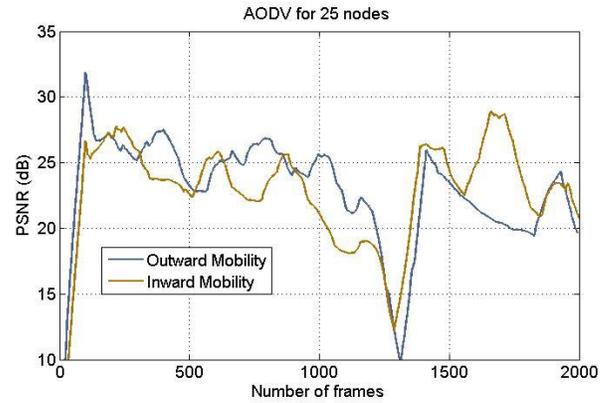

Fig. 14: PSNR performance comparison between inward and outward mobility for AODV

VII. CONCLUSIONS AND PERSPECTIVES

In this paper, we investigated the combined effect of network sparsity and density on PSNR and jitter performances of two MANET routing protocols namely ADOV (reactive) and DSDV (proactive) for video streaming applications. Various network sparsity models were designed with varying network densities. Simulation and evaluation results presented interesting findings. PSNR performance worsens as the density of the network increases. Overall, AODV delivers high levels of PSNR in faster timeframe than DSDV in the various network density and sparsity models analysed in our work. Interestingly, sparsity adversely affects PSNR in a much larger scale than network density for DSDV. In fact, extensive simulations demonstrated that when increasing the network sparsity, DSDV protocol is unable to deliver a video quality that is sufficient enough to extract the PSNR metric.

We also explored the effect two types of mobility on the PSNR metric namely the *Inward* and *Outward Mobility*. It was identified that mobility that results in closely populated networks improves PSNR. Moreover, mobility that results in sparser networks gradually worsens the PSNR performance as the network becomes sparser. Interestingly, in the case of *Outward Mobility*, it was noted that while AODV delivers better PSNR than DSDV initially, DSDV's PSNR outperforms ADOV's as the mobile network becomes sparser.

With respect to jitter performance, results demonstrated that AODV's jitter performance is more resilient to changes in both network density and sparsity than DSDV. However DSDV's jitter performance is much more resilient to the increase in network density than it is to the augmentation in network sparsity.

Finally, it is important to highlight that our work relied on a two-dimensional QoS framework, namely PSNR and jitter given that the latter metrics are of paramount importance in video streaming. For future work, we propose undertaking a multi-dimensional QoS comparative study that encompasses the following QoS metrics: frame loss rate, transit delay, throughput, in addition to PSNR and jitter. Such study will allow a closer analysis of the impact of network density, sparsity, and mobility. We also propose that such a study covers further examples of reactive, proactive and hybrid protocols such as: TORA, OLSR and ZRP.

REFERENCES

[1] S. Ahmad, and J. Reddy, " Delay optimization using Knapsack algorithm for multimedia traffic over MANETs ", in Expert Systems with Applications, 2015, vol 42, no 20, pp. 6819-6827.

[2] A. Jamali, and N. Naja, " Comparative analysis of ad hoc networks routing protocols for multimedia streaming ", in Multimedia Computing and Systems ICMCS'09, IEEE International Conference, 2009, pp. 381-385.

[3] M. Rao, and N. Singh, "Performance analysis of AODV nthBR protocol for multimedia transmission under different traffic conditions for sparse and densely populated MANETs", in IEEE Green Computing and Internet of Things (ICGCIoT), IEEE International Conference, 2015, pp. 1010-1015.

[4] P. Sakalley, J. Kumar, " Review and Analysis of Various Mobile Ad Hoc Network Routing Protocols ", International Journal of Recent Technology and Engineering (IJRTE) ISSN, 2013, pp.2277-3878.

[5] P. K. Bhardwaj, and S. Sharma, and V. Dubey, " Comparative analysis of reactive and proactive protocol of mobile ad-hoc network ", International Journal on Computer Science and Engineering, Vol. 4, No. 7, 2012, pp. 1281.

[6] J. Singh, and U. Goyal, " An Analysis of Ad hoc Routing Protocols ", International Journal of Scientific Engineering and Applied Science (IJSEAS), Vol. 1, No. 2, 2015.

[7] M. Riaz, M. S. I. M. Adnan, and M. Tariqu, " Performance analysis of the Routing protocols for video Streaming over mobile ad hoc Networks ", International Journal of Computer Networks & Communications, Vol. 4, No. 3, 2012, pp. 133-150.

[8] S. Wolf, and M. Pinson, "Video quality measurement techniques", Technical Report 02-392, Department of Commerce, NTIA, USA, 2002.

[9] C. H. Ke, C. K. Shieh, W. Hwang, and A. Ziviani, " An Evaluation Framework for More Realistic Simulations of MPEG Video Transmission ", J. Inf. Sci. Eng, Vol. 24, No. 2, 2008, pp. 425-440.

[10] C. H. Ke, "How to evaluate MPEG video transmission using the NS2 simulator?", in http://csie.nqu.edu.tw/smallko, EE Department, NCKU, Taiwan University, Taiwan, 2006.

[11] J. Klaue, J. Rathke, and A. Wolisz, "Evalvid–A framework for video transmission and quality evaluation", in Computer performance evaluation, Modelling techniques and tools, Springer Berlin Heidelberg, 2003, pp. 255-272.

[12] L. Hanzo, and P. J. Cherriman, Wireless Video Communications, 445 Hoes Lane, Piscataway, 200: IEEE Press, 2001.

[13] K. Stuart, C. Tracy, and C.Michael, " MANET simulation studies: the incredible ", in ACM Mobile Comput Comm Rev, 2005, Vol. 9, No 4.

[14] E. Aguiar, A. Riker, A. Abelém, A. Cerqueira, and M. Mu, " Video quality estimator for wireless mesh networks ", in Quality of Service IEEE 20th International Workshop (IWQoS), 2012, pp. 1-9.

[15] K. Chih-Heng, C. Shieh, W. Hwang, and A. Ziviani," An Evaluation Framework for More Realistic Simulations of MPEG Video Transmission ", J. Inf. Sci. , Vol. 24, No. 2, 2008, pp. 425-440.

[16] I. Agi, and L. Gong, " An empirical study of secure MPEG video transmissions " in Network and Distributed System Security, Proceedings of the IEEE Symposium, 1996, pp. 137-144.

[17] J. Klaue, "EvalVid - A Video Quality Evaluation Tool-set", in http://www2.tkn.tu-berlin.de/research/evalvid/fw.html#bin, Telecommunication and Network Group (TKN), Faculty of EE and CS, Berlin, 2003.

[18] J. Klaue, "EvalVid - A Video Quality Evaluation Tool set, in http://www2.tkn.tu-berlin.de/research/evalvid/cif.html, Telecommunication and Network Group (TKN), Faculty of EE and CS, Berlin, 2003.

[19] A. P. Patil, N.Sambaturu, and K. Chunhaviriyakul, ." "Convergence time evaluation of algorithms in MANETs", in arXiv preprint arXiv:0910.1475, 2009.

AUTHORS PROFILE

**Sabrina Nefti** is currently reading for Masters in Network Architecture at the University of HAL, Batna, Algeria. Certified Information Systems Auditor (UK, 2010). Previous studies and professional experience: MEng in Computer Engineering with First Class Honors from the University of Southampton, United Kingdom of Great Britain (2007), IT Senior Consultant at Ernst & Young (London, UK, 2007-2014), Assistant Engineer at Advanced Risc Machines Limited (Cambridge, UK, 2004-2005); Best Performance Award at MEng degree at Southampton University (UK, 2007), The Coulton Medal for Achievement in Measuring and Control (UK, 2006), Four Zepler Awards for Best Performance at Southampton University, UK (2002, 2003, 2006, 2007). Research interests: 5th generation communication protocols, mobile learning, Internet of Things, pervasive healthcare.

**Maamar Sedrati** received his engineering degree in 1985 from UMC Constantine, Algeria and he obtained the Ph.D. degree in 2011 from UHL Batna University, Algeria. He is currently serving as an assistant professor and member of LaSTIC laboratory at the Computer Science Department, University of Batna 2, Algeria. His research interests include computer networks, Internet technologies and mobile computing, security, quality of service for Multimedia applications in wireless and mobile networks and several aspects of the Internet of Things (IoT). He is a member of technical program committees in many national and international conferences such as ICACIS, CN2TI, IPAC and ICCSA.